# A genome-scale deep learning model to predict gene expression changes of genetic perturbations from multiplex biological networks

Lingmin Zhan[a], Yuanyuan Zhang [a,], Yingdong Wang[a], Aoyi Wang[a], Caiping Cheng[a], Jinzhong Zhao[a], Wuxia Zhang[a, *], Peng Li[a, *], Jianxin Chen[b, *]

[a] College of Basic Sciences, Shanxi Agricultural University, Taigu 030801, China

[b] School of Traditional Chinese Medicine, Beijing University of Chinese Medicine, Beijing 100029, China

*Corresponding author.

E-mail addresses: wxzhang@sxau.edu.cn(W. Zhang), lip@sxau.edu.cn (P. Li), cjx@bucm.edu.cn (J. Chen)

**Abstract**

Systematic characterization of biological effects to genetic perturbation is essential to the application of molecular biology and biomedicine. However, the experimental exhaustion of genetic perturbations on the genome-wide scale is challenging. Here, we show that TranscriptionNet, a deep learning model that integrates multiple biological networks to systematically predict transcriptional profiles to three types of genetic perturbations based on transcriptional profiles induced by genetic perturbations in the L1000 project: RNA interference (RNAi), clustered regularly interspaced short palindromic repeat (CRISPR) and overexpression (OE). TranscriptionNet performs better than existing approaches in predicting inducible gene expression changes for all three types of genetic perturbations. TranscriptionNet can predict transcriptional profiles for all genes in existing biological networks and increases perturbational gene expression changes for each type of genetic perturbation from a few thousand to 26,945 genes. TranscriptionNet demonstrates strong generalization ability when comparing predicted and true gene expression changes on different external tasks. Overall, TranscriptionNet can systemically predict transcriptional consequences induced by perturbing genes on a genome-wide scale and thus holds promise to systemically detect gene function and enhance drug development and target discovery.

**Introduction**

Functional characterization of genes is the core topic of life science, necessary to explore the genetic basis of biological traits, illustrate molecular mechanisms of diseases, and support new drug discovery and development(1-3). In years past, both computational and experimental methods have been used to study gene function. The function of a gene or gene product can be inferred by mapping its sequence in the existing bioinformatic databases. The sequencing of the human genome has revealed a detailed catalog of genetic variation and mutations associated with many diseases(4). However, the known gene-disease associations are insufficient to provide causal or mechanistic insights into uncharacterized genes, further genetic manipulation is required to directly interrogate gene function to understand how genes participate in biological molecular networks and lead to disease states, such as transgenic overexpression (OE), RNA interference (RNAi) and clustered regularly interspaced short palindromic repeat (CRISPR) loss-of-function technologies(5-8). While these operations are often laborious, there is a lack of unified and comprehensive resources for systematically characterizing biological consequences of genetic perturbation at the genome scale. A wonderful solution is to create a catalog of cellular signatures representing the overall effects of perturbation of all genes in the genome. Following this concept, Connectivity Map (CMap) is developed as a public compendium of transcriptional responses of genetic perturbation and has curates ~400 000 gene expression profiles induced by three types of genetic perturbations on various cell lines, including OE, RNAi, and CRISPR(9,10). This dataset provides opportunities to build functional connections between drugs, genes, and diseases at a gene expression level. Despite this, its small scale limits its utility. The newest CMap (updated in December 2020) contains genetic perturbations for only thousands of genes corresponding to OE, RNAi, and CRISPR respectively, which is much less than the number of genes (>20 000) across the whole human genome. This situation prompts us to establish a model for predicting inducible gene expression changes (GECs) by perturbing every gene in the genome and this genome-scale resource can accelerate the characterization of gene function from a systematic level.

It has been known that different genes with similar features in biological networks function similarly(11,12). Therefore, in this work, we first produce integrated network features from various biological networks for each gene in the human genome. Then we build a neural network model to map specific gene network features to GECs induced by corresponding genetic perturbations. In addition, we consider there may be complementary information between GECs of three types of genetic perturbations for the same genes(13), we improve the predicted GECs for one type of genetic perturbation by integrating the other two GECs for the same genes by a self-attention architecture.

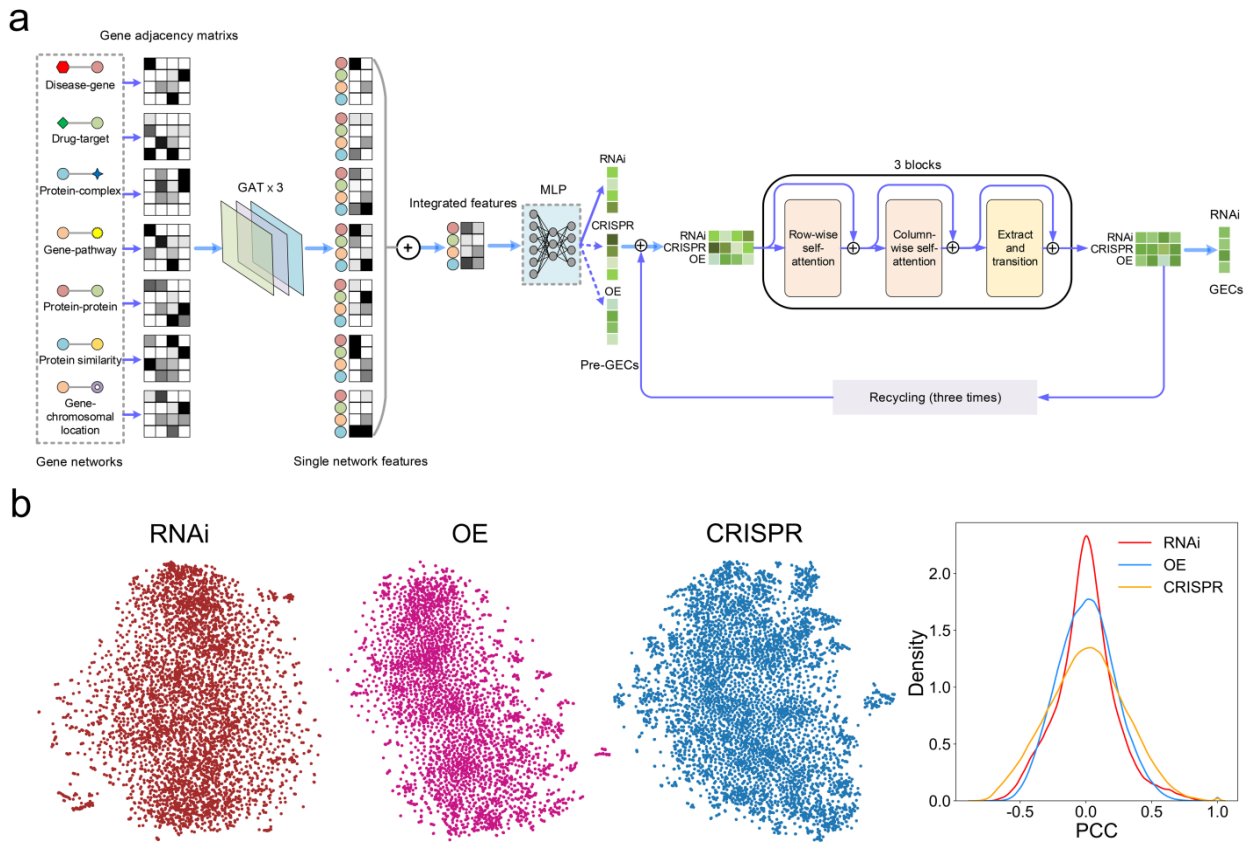

**Figure 1 | TranscriptionNet framework and distribution of GECs for RNAi, OE, and CRISPR.**
**a.** The architecture of the TranscriptionNet model to predict gene expression changes (GECs) of RNAi. TranscriptionNet uses two-stage networks of FunDNN (Functional network-based Deep Neural Network) and GenSAN (Genetic perturbation-based Self-Attention Network) to load the genome-wide functional connection knowledge among genes and complementary information between different types of genetic interference manners on the same genes, respectively. FunDNN uses adjacency matrices for multiplex gene interaction networks as input. Each network passes through a Graph Attention Network (GAT) to generate network-specific gene features, which are then combined into the integrated features. A stack of three GAT layers are used to generate gene features encompassing larger neighborhoods. The integrated features are propagated through multiple dense layers to fit pre-GECs induced by each type of genetic perturbation (RNAi, OE, and CRISPR). GenSAN takes GECs for all three types of genetic perturbations as input. To predict GECs for one type of genetic perturbation (e.g. RNAi), GenSAN processes the pre-GECs of RNAi and true GECs of CRISPR and OE together by a multi-layer transformer encoder block (with axial self-attention in a dual-tower architecture) to capture complementary gene expression information between them. In the absence of known GECs for CRISPR or OE, the corresponding pre-GECs are used as alternatives. Moreover, the self-attention block is reinforced by feeding the processed GECs recursively into the same modules (named "recycling"). Finally, the predicted GECs for RNAi can be extracted from the output GEC matrix. **b.** t-distributed stochastic neighbor embedding (t-SNE) analysis is performed on the GECs of the three genetic perturbations (RNAi, OE, and CRISPR), demonstrating the uniformity of data distribution. The distribution of Pearson correlation coefficients among GECs of each genetic perturbation reveals the differences among different GECs.

## Results

**Problem formulation**

Our task is to build a model to systemically predict transcriptional consequences induced by perturbing genes on a genome-wide scale. The transcriptional consequences adopt gene expression changes (GECs) of 978 landmark genes in the L1000 project, which has been verified to represent the reduced representation of cell transcriptome and can be used to infer the remainder of the transcriptional profile(10). The genetic perturbations correspond to three classic genetic techniques, RNAi, CRISPR, and OE. Our first hypothesis is that the interference on different genes with similar biological functions causes similar biological consequences, including transcriptional profiles. Functional associations among genes are often represented as various gene-gene networks(14). We can use these networks to learn the network-specific feature representation for each gene. The model is designed to map the gene representation to its GECs. The other assumption is that GECs induced by three types of genetic perturbations for the same gene have part similar patterns in their inducible gene expression profiles. Thus, when predicting GECs of one type of genetic perturbation, the other two types of perturbations can supply complementary information for expression genes.

**TranscriptionNet architecture and training**

The proposed TranscriptionNet adopts a two-stage coarse-to-fine network architecture, as shown in **Figure 1a**. This coarse-to-fine network framework has been successfully used in image fields such as image inpainting and deblurring(15,16), to improve the accuracy and generalizability of image processing. Inspired by this notion, we here take advantage of the two-stage networks to load the genome-wide functional connection knowledge among genes and complementary information between different types of genetic interference manners on the same genes, respectively. The first network that we name FunDNN (**Fun**ctional network-based **D**eep **N**eural **N**etwork) is devoted to making gene functional representations and fitting the gene representation to its GECs induced by perturbing this gene. Specifically, FunDNN integrates various large-scale gene functional networks to encode a unified representation for each gene through a deep learning-based network integration algorithm, BIONIC (Biological Network Integration using Convolutions), which has been proved to perform better than existing network embedding methods on a range of benchmark tasks(17). Each network-specific representation is then run through a sequence of fully connected layers to learn the first stage GECs (termed pre-GECs) for each type of genetic perturbation. The second network that we term GenSAN (**Gen**etic perturbation type-based **S**elf-**A**ttention **N**etwork) takes GECs for all three types of genetic perturbations as input. To predict GECs for one type of genetic perturbation (e.g. RNAi), GenSAN processes the pre-GECs of RNAi and true GECs of CRISPR and OE together by a multi-layer transformer encoder block to capture complementary gene expression information between them. In the absence of known GECs for CRISPR or OE, the corresponding pre-GECs are used as alternatives. Moreover, the self-attention block is reinforced by feeding the processed GECs recursively into the same modules (named "recycling"). The pre-GECs internalizing complementary information from the other two types of genetic perturbations pass multiple dense layers to obtain final predictive GECs.

We use the latest high-throughput CMAP dataset from the L1000 platform to fit the model. The dataset contains GECs induced by three types of genetic perturbations on 8184 genes, corresponding to 4454 RNAi, 5139 CRISPR, and 3538 OE. Values of each gene are normalized with the MinMax scaler (see methods). Analysis of GECs for each type of genetic perturbation using t-distributed stochastic neighbor embedding (t-SNE) shows that all types of data are distributed uniformly (**Figure. 1b**). As the big difference between GECs induced by the three types of genetic manipulation techniques(10,13), each type is trained independently with similar model architecture but different hyperparameters. The input of FunDNN includes seven diverse human gene networks: the disease-based gene association network (995 genes, 4,047 interactions), the drug-based gene association network (2,792 genes, 131,193 interactions), the protein complex-based network (3,407 genes, 40,170 interactions), the pathway-based gene network (10,623 genes, 178,7207 interactions), the chromosomal location-based gene network (26,813 genes, 860,164 interactions), the STRING protein-protein interaction network (17,844 genes, 535,462 interactions) and the protein sequence similarity network (18,586 genes, 415,6924 interactions), which combine for a total of 26,945 unique genes and 751,5167 unique interactions (Supplementary table 1). For GenSAN, it receives the pre-GECs for one type of genetic perturbation from the upstream FunDNN, combined with true GECs (or corresponding pre-GECs as alternatives in the absence of true GECs) for the other two types of genetic perturbations on the same gene. Both FunDNN and GenSAN use a customized reconstruction loss termed PMSE that combines the Pearson correlation and Mean Squared Error (MSE) to minimize the discrepancy between predicted and true GECs.

**Evaluation strategies and metrics**

For each type of genetic perturbation, all data are randomly split into training, validation, and test sets with a 7:1:2 ratio. The training set is used to fit the model, whose performance is evaluated by the hold-out test set. Pearson correlation coefficients (PCC) are used as the major metric to evaluate the performance of models. We can quantify the correlation for each pair of predicted and true GECs and compare different models through the distribution or average of PCC values in the test set. In addition, MSE is used to assess the numerical difference between predicted and true GECs, and the Kolmogorov-Smirnov (KS) test statistic maximum distance (D) is used to compare the coherence of distribution of predicted and true GECs.

To further evaluate the generalization ability of TranscriptionNet, we predict GECs by TranscriptionNet for external genetic perturbations (corresponding to all network genes whose GECs are not profiled in the L1000 project) and compare their quality with known GECs in the L1000 project through the following analyses: (1) gene coannotation prediction; (2) compound-target interaction prediction and (3) disease-gene association detection. First, we compare the performance of external and known GECs to identify gene pairs coannotated to the same functional term using a binary classification strategy in which gene pairs within at least one functional annotation are regarded as positive pairs, while gene pairs not within an annotation are retained as negative pairs (See methods)(17). Second, we compare the ability of external and known GECs to recover known compound-target interactions based simply on the similarity between GECs of genetic perturbations and compounds of targets. Finally, we compared the quality of external and known GECs, using the same binary classification strategy to identify genes associated with disease. The known disease-associated genes are considered to be positive, while genes not in the disease-associated gene set were retained as negative (See methods). These external evaluations determine how effectively the predicted GECs can be used for additional tasks compared with existing GECs.

**Performance evaluation**

Prediction of GECs is a classical multivariate linear regression (MLR) problem in machine learning (ML). Here, we compare TranscriptionNet to four baseline regression algorithms under the random split setting: Decision Tree Regression (DTR), K-Nearest Neighbors Regression (KNR), Linear Regression (LR), and Random Forest Regression (RFR). The inputs of the four regressors are network-specific gene features derived from BIONIC, similar to FunDNN that can be regarded as a deep neural network regression algorithm. As shown in **Figure 2a**, the GECs predicted by these regression algorithms are all well-fitted with true GECs in the test set (PCC averages of 0.749 to 0.865 for RNAi, 0.754 to 0.870 for CRISPR (Supplementary Figure 1) and 0.875 to 0.935 for OE (Supplementary Figure 2)), convincing the reliability of mapping robust gene functional representations onto transcriptional responses of the corresponding genetic perturbations. Especially, for all three types of genetic perturbations, FunDNN performs better than the four ML regressors in terms of PCC averages, while its performance in MSE and D is also competitive. Furthermore, TranscriptionNet which combines FunDNN and GenSAN, consistently outperforms baselines and individual FunDNN, with optimal PCC and D averages for all three types of genetic perturbations. We also compare the detailed distribution of all metrics between different models. TranscriptionNet also outperforms baselines and FunDNN for almost all of GECs (**Figure 2b**). The profile-wise comparative analysis shows that the percentage of GECs predicted by TranscriptionNet with larger PCC outperforms FunDNN, DTR, KNR, LR, and RFR by 87.42%, 100.00%, 94.61%, 94.16%, and 51.24%, respectively (**Figure 2c**). These results indicate that by importing correlation information between three types of genetic perturbations on the same gene, GenSAN can effectively improve the quality of pre-GECs predicted by FunDNN.

Network integration is the primary module for TranscriptionNet. An excellent network integration algorithm should produce accurate and comprehensive gene representations from biological networks. BIONIC used in our model has been proven to outperform existing integration methods across all evaluation types and benchmarks(17). To further confirm the potency of BIONIC in our experiment, we replace BIONIC in TranscriptionNet with three different established integration approaches to compare their performance: a naive union of networks (Union), a deep learning multi-modal autoencoder (deepNF)(18), and a multi-network extension of the node2vec(19) model (multi-node2vec)(20). We observe that BIONIC outperforms the established integration approaches in terms of PCC, MSE, and D (Supplementary Figure 3). In addition, to ensure the advantage of gene features encoded from multiple networks over single networks, we compare the results of TranscriptionNet using multiple networks with those using single networks. As expected, gene features learned from multiple networks perform as well as, or better than the individual input networks across all evaluation metrics (see the results for RNAi, OE, and CRISPR in Supplementary Figure 4, 5, and 6, respectively). These results demonstrate the strength of BIONIC for encoding suitable gene representations in our model.

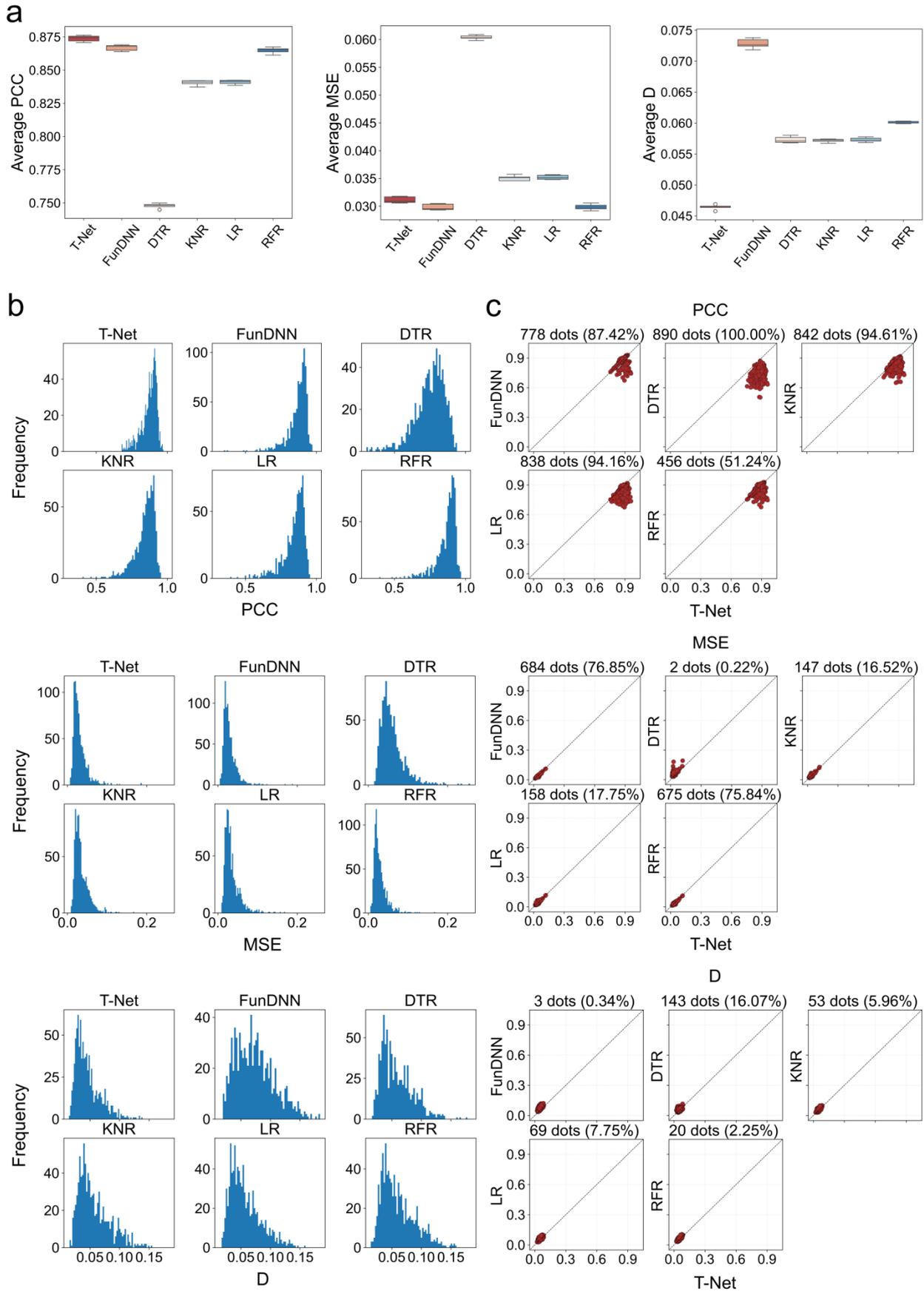

**Figure 2 | Comparison of TranscriptionNet to baseline models for predicting GECs of RNAi. a.** Box plots for

three metrics (Pearson correlation coefficients (PCC), Mean Square Error (MSE), Kolmogorov-Smirnov (KS) test statistic maximum distance (D)) on the test dataset for six models: TranscriptionNet, FunDNN, DTR, KNR, LR, and RFR. **b.** Distribution of PCC, MSE, and D for the six models on the test dataset. Data are obtained from five random runs. **c.** The profile-wise comparative analysis of PCC, MSE, and D for each predicted GECs in the test set between TranscriptionNet and the other five models. The x-axis represents the results predicted by the TranscriptionNet model, while the y-axis represents the results predicted by the other five models. The dots below the diagonal indicate that TranscriptionNet has higher PCC values and lower MSE and D values compared to other models.

**Characterization of gene function**

TranscriptionNet is used to predict GECs for external genetic perturbations corresponding to gene members in all input networks except perturbations profiled in the CMap dataset, resulting GECs for 22,496 RNAi, 21,806 CRISPR and 23,427 OE (https://github.com/lipi12q/TranscriptionNet/tree/master). To assess the quality of these predicted GECs, we compare the ability of predicted and known GECs to recover gene pairs coannotated to the same functional term. For all three types of genetic perturbations, the external GECs have similar performance to known GECs at identifying coannotated gene pairs with over two different functional benchmarks that are not used in our model: Kyoto Encyclopedia of Genes and Genomes (KEGG) pathways(21) and GO Biological Processes (BP)(22) (**Figure 3**). The similar performance between external and known GECs is obtained for coannotated analyses in the functional networks used in our model, including the disease-based gene association network, the drug-based gene association network, the protein complex-based network, the pathway-based gene network, the chromosomal location-based gene network, the STRING protein-protein interaction network and the protein sequence similarity network(Supplementary Figure 7). The similar performance of external and known GECs in the characterization of gene function verifies the high quality of GECs predicted by TranscriptionNet.

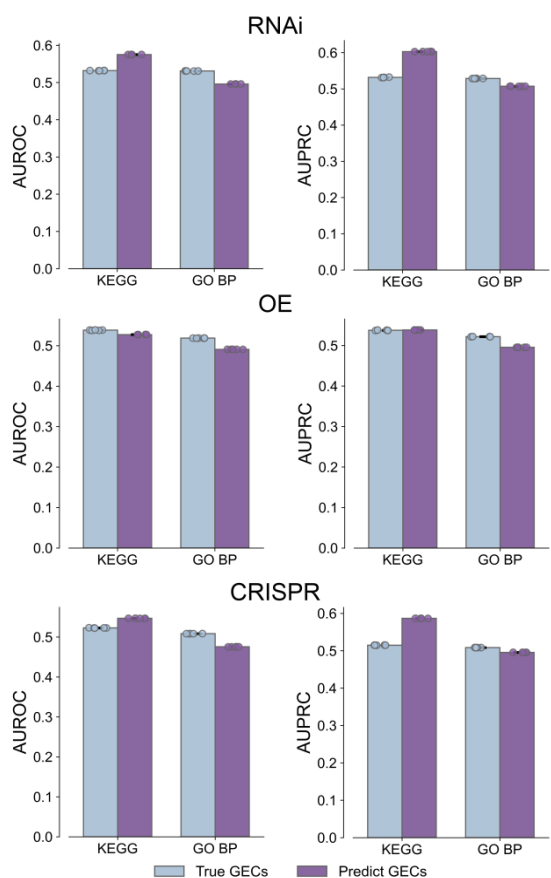

**Figure 3 | Gene function characterization by true and predicted GECs.** The co-annotation prediction evaluations of true and predicted GECs in Kyoto Encyclopedia of Genes and Genomes (KEGG) pathways and GO Biological Processes (BP) benchmarks using RNAi, OE, and CRISPR genetic perturbations. The statistics are carried out over five random runs. The evaluation criteria include the area under the receiver operating characteristic curve (AUROC) and the area under the precision-recall curve (AUPRC). The data are represented as the average value, with error bars indicating the 95% confidence interval of 5 independent samples, and floating points representing the accurate values of 5 independent samples.

**Characterization of compound-target interactions**

Theoretically, we can interrogate connections between drugs and protein targets simply through correlations of transcriptional profiles induced by drugs and genetic perturbations. To assess the feasibility of using predicted GECs in such an approach, we compare the ability to recover known drug-target interactions by predicted GECs and known GECs. We directly calculate Pearson correlation coefficients between 33,609 compounds and 26,945 genetic perturbations (including known GECs and predicted GECs), resulting in 905,459,780 relationships. Among them, 8,995 pairs have been curated in CMap as real drug-target interactions (Supplementary Data 1). We refer to these interactions as the positive set, while all remaining pairs are retained as the negative set. Firstly, for each type of genetic perturbation, we compare the distribution of correlations between drugs and targets with known and predicted GECs in positive and negative sets, respectively. As shown in **Figure 4a**, we observe that whether for predicted or true GECs, drug-target pairs in the positive set are extremely more correlated than those in the negative set. Moreover, there is a similar distribution of drug-target correlations for true and predicted GECs in positive or negative sets.

We further quantify the performance of GECs for discriminating different types of drug-target pairs by a receiver operator characteristic (ROC) curve. We calculate the true positive rate and the false positive rate and plot ROC curves for known and predicted GECs based on various thresholds of correlation coefficients. As shown in **Figure 4b**, we find both known and predicted GECs perform similarly and well for all three types of genetic perturbations (AUROC of 0.74 and 0.76 for RNAi, 0.80 and 0.82 for CRISPR and all 0.68 for OE). When correlations are summarized across all three types of perturbations by the logistic linear regression algorithm, the performance has a minor improvement with AUROC of 0.81 and 0.85 for known and predicted GECs, respectively (**Figure 4c**).

These results indicate the GECs predicted by TranscriptionNet are similar to known GECs and can be used to explore drug-target interactions.

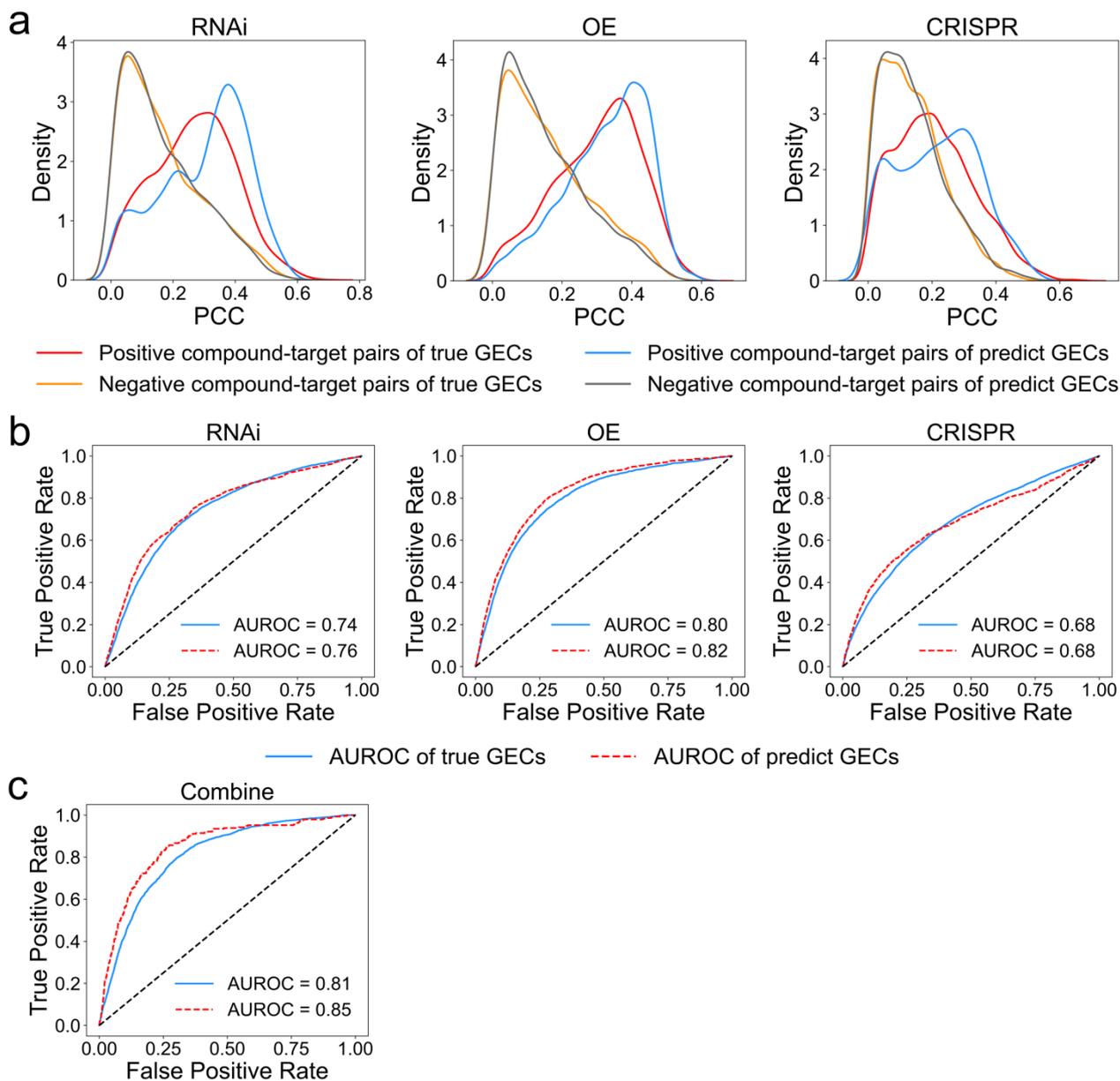

**Figure 4 | Characterization of compound-target interactions by true and predicted GECs. a.** The distribution of PCC between GECs in positive and negative compound-target pairs for true and predicted GECs. The distribution curves are separately plotted for RNAi, OE, and CRISPR perturbations. **b.** ROC curves for predicting compound-target interactions by true and predicted GECs correlation for the three genetic perturbations: RNAi, OE, and CRISPR. **c.** ROC curve of true and predicted GECs correlation by combining three types of genetic perturbations for the same genes using a logistic regression algorithm.

**Characterization of disease-gene associations**

To evaluate the feasibility of using predicted GECs to study the association between diseases and target genes, we compared the ability of predicted GECs and known GECs to recover known disease-gene associations. Based on the differential gene expression profiles induced by ischemic cardiomyopathy and non-ischemic cardiomyopathy, we directly calculated the PCC between GECs for all 26,945 genetic perturbations (including known GECs and predicted GECs) and the differential expression profiles of the two diseases. The resulting gene lists ranked by PCC are separately mapped to known 110 and 15 genes associated with ischemic cardiomyopathy and non-ischemic cardiomyopathy. Among them, the genes associated with cardiomyopathy are referred to as the positive set, while all other genes are retained as the negative set. The performance of GECs in distinguishing different types of disease-gene pairs is further quantified by AUROC and AUPRC. We find that true and predicted GECs perform similarly across three genetic perturbations (**Figure 5**). These results demonstrate that the GECs predicted by TranscriptionNet have a similar ability to known GECs to characterize disease-gene associations, confirming the quality of the GECs predicted by TranscriptionNet.

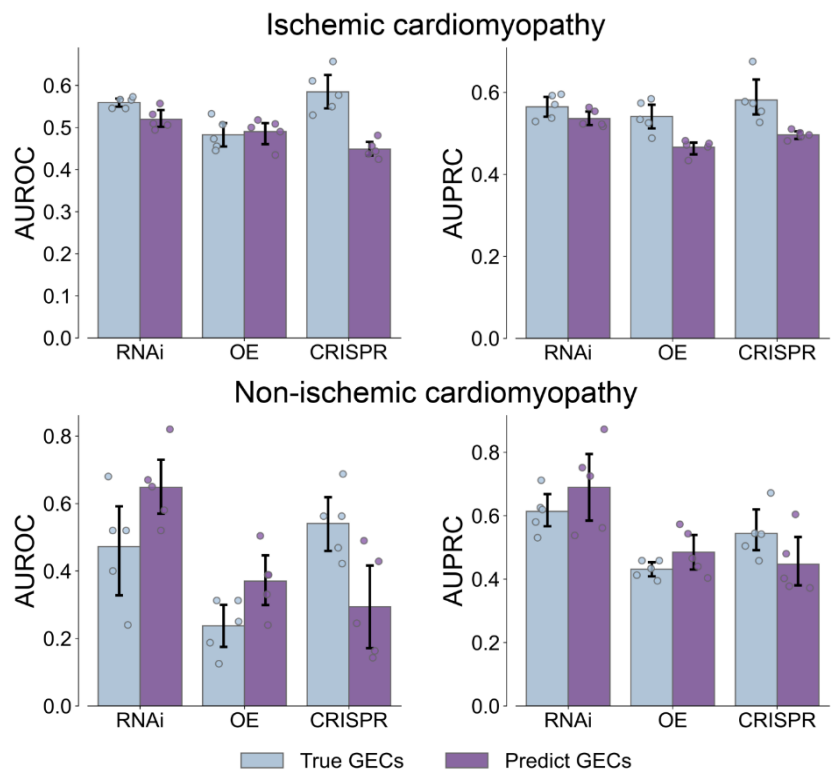

**Figure 5 | Characterizing associated genes for ischemic cardiomyopathy and non-ischemic cardiomyopathy by true and predicted GECs**. The performance of predicting disease-gene

associations by directly calculating PCC between GECs induced by genes and GECs induced by ischemic cardiomyopathy and non-ischemic cardiomyopathy. We separately analyze the three types of genetic perturbations: RNAi, OE, and CRISPR. The statistics are carried out over five random runs. The evaluation criteria include the area under the receiver operating characteristic curve (AUROC) and the area under the precision-recall curve (AUPRC). The data are represented as the average value, with error bars indicating the 95% confidence interval of 5 independent samples, and floating points representing the accurate values of 5 independent samples.

**Discussion**

Systematic characterization of gene function can enhance the exploration of pathological mechanisms of diseases and aid in target-gene-based drug discovery. As transcriptional profiles represent the overall molecular activities of cells, genetic perturbation-induced transcriptional profiles are excellent molecular features to characterize the biological function of genes. Although techniques like RNA-array and high-throughput sequencing have been extremely applied for gene expression profiling, their genome-wide scalability is limited due to huge costs. TranscriptionNet can serve as a valuable complement to these experiments. We train TranscriptionNet based on transcriptional profiles associated with three different perturbations (RNAi, CRISPR, and OE) on thousands of genes that are generated by the L1000 project(10). TranscriptionNet has been used to infer GECs of unknown genetic perturbations on the genome-wide scale and increases perturbational GECs from thousands of genes to 26,945 genes for each type of genetic perturbation. Moreover, the comparison between these predicted and known GECs on different external tasks demonstrates the generalization ability of TranscriptionNet.

For reliable predictions, TranscriptionNet must be trained based on multiplex biological networks as most networks have no uniform scale and quality, and one network generally only focuses on certain types of functional relationships. In addition, the complementary effects between the GECs by RNAi, CRISPR, and OE for the same genes are based on the assumption that different types of genetic manipulations on the same genes theoretically should produce similar biological outcomes. However, the fact is that these genetic perturbations are inhomogeneous due to various confounding factors, such as systematic and off-target effects, non-specific toxicity, and different mechanisms of action(10,13,23).

One of the most important strengths of TranscriptionNet is detecting biological connections between drugs, genes, and diseases. Based on the genome-wide scale transcriptional profiles generated by TranscriptionNet, we can explore biological targets and pathways for drugs and genes associated with pathogenic mechanisms of diseases by directly comparing transcriptional profiles of genetic perturbations and existing high-throughput data, and further identify candidate disease treatments with specific molecular mechanisms. Therefore, TranscriptionNet holds promise to not only systemically detect gene function but also enhance drug development and target discovery.

## Methods

### Network preprocessing

To incorporate as much gene functional association information as possible, seven classes of gene interaction networks are curated from different biological repositories. The disease-based gene association network is constructed by connecting two genes related to the same diseases based on the Online Mendelian Inheritance in Man (OMIM), a comprehensive knowledgebase that covers human genes and genetic disorders relationships (https://www.omim.org/)(24). The drug-based gene association network is built by linking two proteins targeted by the same drugs based on known drug-target interactions, which are downloaded from the DrugBank database (https://www.drugbank.ca/). The protein complex-based network is created by connecting two genes in the same complex subunits, which are collected from CORUM, a database that provides a manually curated catalog of experimentally characterized protein complexes from mammalian organisms (https://mips.helmholtz-muenchen.de/corum/)(25). The pathway-based gene network is curated by linking two genes in the same biological pathways, which are downloaded from Reactome, a knowledgebase of biological pathways, reactions, proteins, and molecules (https://reactome.org)(26). The chromosomal location-based gene network is built by connecting two genes in the same cytogenetic bands, which are curated from Gene, a searchable database of gene-specific contents in the National Center for Biotechnology Information (NCBI) (http://www.ncbi.nlm.nih.gov/gene). The five gene networks are constructed based on the assumption that two genes associated with the same biological entities should be more functionally related than two genes associated with different biological entities. The similarity between two genes in these networks is quantified by calculating Jaccard similarity scores. The Search Tool for Recurring Instances of Neighboring Genes (STRING; https://string-db.org) quantitatively integrates different studies and interaction types into a single integrated score for each gene pair based on the total weight of evidence(27). The protein sequence similarity network is obtained by calculating pairwise Smith–Waterman scores(28). The sequences of reviewed human proteins are collected from UniProt (https://www.uniprot.org/)(29). To obtain networks that are comparable in size to other networks, the STRING network is filtered for only the top 10% of interactions by interaction scores, the pathway network keeps gene pairs with a similarity magnitude greater than or equal to 0.2 as edges, and the sequence similarity network retains edges with pairwise similarity scores greater than or equal to 0.23. To unify all networks for analysis, gene names in each network are mapped to human Entrez gene ID using the R package org.Hs.eg.db (version:3.18.0). All these networks cover 26,945 unique human genes and the detailed information of each network is provided in the Supplementary Table 1.

**Gene expression data source and preprocessing**

To build the CMAP dataset, the L1000 platform directly measures 978 landmark genes using Luminex bead-based technology and infers an additional 11,350 genes, of which, 9,196 are well inferred(10). We download the LEVEL 5 data from the Expanded CMap LINCS Resource 2020 (version: beta) at the LINCS data releases app (https://clue.io/releases/data-dashboard). We focus on the 978 landmark genes of three types of genetic perturbations, including 23,835 shRNA, 142,901 CRISPR, and 34,171 OE treatments. We collate gene targets of genetic perturbagens from the metadata and check the names of all target genes using the R package HGNChelper (version: 0.8.1). In each type of genetic perturbation, we combine all transcriptional profiles for each target gene by the weighted average algorithm, which calculates a weighted average of the gene expression signatures of each target, with coefficients given by a pairwise Spearman correlation matrix between the expression profiles of all signatures(13). Finally, we obtain unique transcriptional profiles for 8288 genes, corresponding to 4,454 shRNA, 5,319 CRISPR, and 3,538 OE target genes. For each gene in all profiles, we normalize them to [-1, 1] using the MinMax Scaler in Python's scikit-learn (version: 1.3.2), to reduce the differences between genes and accelerate model convergence. In the prediction model of transcriptional profiles, data in each type of genetic perturbation is randomly divided into training, validation, and test sets with a 7:1:2 ratio.

**TranscriptionNet architecture**

TranscriptionNet comprises two stages. The first network that we term FunDNN (**Fun**ctional network based **D**eep **N**eural **N**etwork) takes an array of gene functional networks as inputs to produce pre-transcriptional profiles for RNAi, CRISPR, or OE perturbations. FunDNN contains two main blocks. The first block processes heterogeneous gene networks through a sequence of graph attention network (GAT) layers to learn a unified representation for each gene. Then the integrated gene features are fed into a multi-layer neural network to generate pre-transcriptional profiles for each genetic perturbation. The second network that we term GenSAN (**Gen**etic perturbation type-based **S**elf-**A**ttention **N**etwork) processes pre-transcriptional profiles by axial self-attention framework to capture complementary gene expression information for one type of genetic perturbation from the other two types.

**Functional network-based deep neural network**

Robust and integrated gene representations are learned from various functional networks using the general and scalable deep learning framework for network integration termed BIONIC (Biological Network Integration using Convolutions). This architecture is selected as its encoded features contain substantially more topological and functional information compared to existing architectures(17). Specifically, each input gene network is represented by its adjacency matrix A where $A_{ij} = A_{ji}$ =edge weight value if node *i* and node *j* share an edge and $A_{ij} = A_{ji} = 0$ otherwise. BIONIC encodes each input network using three sequential graph attention network (GAT)(30) layers. The gene encoder is described as follows:

$$\text{GAT}(A, H) = \sigma(HW^T) \quad (1)$$

Where

$$a_{ij} = \frac{A_{ij}\exp(\sigma(\partial^T[Wh_i||Wh_j]))}{\sum_{K=1} A_{ij}\exp(\sigma(\partial^T[Wh_i||Wh_j]))} \quad (2)$$

Here, W is the layer-specific trainable weight matrix. $\partial$ is the vector of learnable attention coefficients. K corresponds to nodes in the neighborhood of i. $h_i$ is the feature vector of node i, that is, the *i*th row of feature matrix H. The initial feature matrix $H_{init}$ is an identity matrix so that each node is uniquely identified. These node features are further mapped to a real-valued dense matrix with a dimension of 2048 through a learned linear transformation. σ represents the nonlinear function LeakyReLU. In each GAT layer, the multi-head attention scheme is learned as:

$$\text{GAT}(A, H) = ||_{k=1}^{K} \sigma\left(a^{(k)} H W^{(k)T}\right) \quad (3)$$

where The number of heads K=10. After each network is encoded, the network-specific node features are then combined to produce the final unfiled features through a weighted, stochastically masked summation as follows:

$$H_{combined} = \sum_{j=1}^{N} S_j m^{(j)} \odot H^{(j)} \quad (4)$$

Here, N is the number of input networks. $S_j$ is the learned scaling coefficient for feature representations of network j, which enables BIONIC to scale features in a network-wise fashion. All values in S should be positive and sum to 1. $m^{(j)}$ is the node-wise stochastic mask for network j, which is designed to randomly drop node feature vectors produced from some networks, forcing the network encoders to learn cross-network dependencies. $\odot$ is the element-wise product and $H^{(j)}$ is the learned feature matrix for nodes network j.

BIONIC maps $H_{combined}$ to a low-dimensional feature matrix F with a dimension of 512 through a learned linear transformation. In F, each row corresponds to a node feature. BIONIC can decode F into reconstructions of the original input networks. To obtain a high-quality F, BIONIC uses an unsupervised training objective to minimize the gap between the reconstructed and the input networks:

$$L_{unsupervised} = \frac{1}{n^2} \sum_{j=1}^{N} ||b^{(j)} \odot (\hat{A} - A^{(j)}) \odot b^{(j)}||_F^2 \quad (5)$$

Where the reconstructed network $\hat{A} = F \cdot F^T$, n is the total number of nodes present in the union of networks, $b^{(j)}$ is a binary mask vector for network j indicating which nodes are present (value of 1) or extended (value of 0) in the network, $A^{(j)}$ is the adjacency matrix for network j and $||\cdot||_F$ is the Frobenius norm. This loss represents computing the mean squared error between the reconstructed network $\hat{A}$ and input $A^{(j)}$ while the mask vectors remove the penalty for reconstructing nodes that are not in the original network j (that is, extended), then summing the error for all networks.

The integrated network features for each gene are further fed into the second block of FunDNN, a multi-layer perceptron (MLP). The MLP model uses multiple dense layers to capture information between different features for each gene described as follows:

$$F^{l+1} = (\sigma(W^l F^l + b^l)) W^o + b^o$$

Where σ is the non-linear activation function, $W^l$ and $b^l$ are model weights and bias at the l-th layer, $W^o$ and $b^o$ are model weights and biases at the output layer. Each hidden layer has the same number of nodes while following a dropout layer. The activation function of all layers is LeakyReLu except the penultimate layer is Tanh. The role of the Tanh function is to compress the features in the range of [−1, 1]. The output layer is an innocent linear layer that maps gene features to the predicted transcriptional profiles $\hat{t}$ with a dimension of 978 in the range [−∞, +∞]. To predict transcriptional profiles of genetic perturbations with less discrepancy in the expression of corresponding genes to its true transcriptional profiles, we design a customized loss function PMSE that is a weighted sum of the mean squared error (MSE) and Pearson correlation losses:

$$L_{MSE} = \frac{1}{n}\sum_{i=1}^{n}(t_i - \hat{t}_i)^2 \quad (1)$$

$$L_{Pearson} = \frac{\sum_{i=1}^{n}(t_i - \bar{t})(\hat{t}_i - \bar{\hat{t}})}{\sqrt{\sum_{i=1}^{n}(t_i - \bar{t})^2}\sqrt{\sum_{i=1}^{n}(\hat{t}_i - \bar{\hat{t}})^2}} \quad (2)$$

$$PMSE = (1 - \beta)L_{MSE} + \beta L_{Pearson} \quad (3)$$

Where $t$ is the true GECs of the corresponding target genes, $\bar{t}$ and $\bar{\hat{t}}$ are the average values of the true and predicted GECs, respectively. β is a hyperparameter in the range [0, 1] indicating the relative weight of the two losses. To predict transcriptional profiles for RNAi, CRISPR, or OE perturbations, all weights in the MLP model are separately updated for each type by backpropagation using the Adadelta optimizer. Hyperparameter combinations are automatically chosen for each type of perturbation using Optuna (version: 3.2)(31) and provided in Supplementary Data 2.

**Genetic perturbation type-based self-attention**

Based on the transcriptional profiles predicted by FunDNN (pre-GECs), the second network GenSAN intends to capture complementary information between transcriptional profiles of the three types of genetic perturbations, RNAi, CRISPR, and OE, as well as complementary information between 978 marker genes, through an axial self-attention[32] framework to refine pre-GECs. For example, to refine pre-GECs of RNAi, true GECs of the other two types of perturbations CRISPR and OE for the same gene are loaded together as inputs. The input matrix $\hat{T}_{init}$ has a dimension of 3 × 978. The first row is the pre-GECs of RNAi, and the other two rows are true GECs of CRISPR and OE for the same gene. If lacking true GECs of CRISPR or OE, the corresponding pre-GECs are used.

We use multiple transformer encoder units(33) to effectively learn the interrelated information between transcriptional profiles of the three types of genetic perturbations and 978 marker genes:

$$\hat{T}_r^l = rowatten(\hat{T}^l, W_r^l, b_r^l) \quad (1)$$

$$\hat{T}_c^l = colatten((\hat{T}_r^l)', W_c^l, b_c^l) \quad (2)$$

$$\hat{T}^{l+1} = \left(\sigma(W_f^l \hat{T}_c^l + b_f^l)\right) W^o + b^o \quad (3)$$

Where each layer $l$ corresponds to a transformer encoder unit and consists of an axial self-attention layer and a feed-forward neural network layer. The row-wise self-attention block establishes attention weights for the transcriptional profiles of the three genetic perturbations, capturing complementary information between them. The column-wise self-attention block allows for the exchange of information between the 978 marker genes. $W_r^l$ and $b_r^l$, $W_c^l$ and $b_c^l$, $W_f^l$ and $b_f^l$ are learnable weight matrices and bias vectors of the row-wise self-attention block, column-wise self-attention block, and feed-forward neural network layer in the $l$th transformer encoder unit. $W^o$ and $b^o$ are the weights and bias at the output layer of the feed-forward neural network. $\hat{T}^l$ is the $l$th hidden transcriptional profile and $\hat{T}^0 = \hat{T}_{init}$. $\hat{T}_r^l$, $\hat{T}_c^l$ are the hidden transcriptional profiles of the $l$th row-wise self-attention block and column-wise self-attention block, respectively. $(\hat{T}_r^l)'$ is the transposed transcription profiles of $\hat{T}_r^l$, with a dimension of 978×3. σ is the activation function LeakyReLU. Moreover, an iterative refinement termed "recycling" is applied to the attention stack, the resulting matrix is recycled and iteratively updates the input data. Each recycling combines inputs and outputs from the last iteration and produces reinforced outputs with shared weights. The recycling process creates a recurrent network and deepens the whole network without significantly increasing training time and the number of parameters. This has been successfully applied in other areas such as computer vision and protein structure prediction(34,35).

After processing with attention stacks, the representation of the specific perturbation is extracted from the first row in the output matrix and processed by a multi-layer dense network to predict the transcriptional profile t̂. Similar to the prediction of FunDNN, the loss function PMSE is used to train the model.

We use the stochastic gradient descent optimization algorithm to train the model. The hyperparameters of the three types of perturbation models are mostly the same except for the batch size and initial learning rate (Supplementary Table 2). We use Adam with a weight decay of 1e-5, a learning rate warm-up of the first 5 epochs, a linear increase in the learning rate during the prediction phase, and a slow decrease according to the cosine function. After 110 epochs, we stop learning. We use a dropout probability of 0.05 on all layers, and the number of nodes in the hidden layer of the feedforward neural network in each Transformer unit is set to 1024. To avoid neuron death, we use the LeakyReLU activation function instead of the standard ReLu. The learning rate, batch size, number of transformer unit layers, number of attention heads, and number of cycles are manually adjusted throughout the training process, following the tuning method of comparing the three indicators of PCC, MSE, and D on the test set. When the three indicators stop improving, we consider the current hyperparameter combination to be the optimal combination for the model.

All models were trained on an NVIDIA A100 graphics processing unit with 80GB of graphics memory, 128 GB of system memory, and 13 Intel® Xeon® CPUs running at 2.10 GHz.

**Baselines**

The task of predicting transcriptional profiles can be regarded as a multivariate linear regression (MLR) problem. Therefore, the performance of TranscriptionNet is first compared with that of classical MLR models Decision Tree Regression (DTR), K-Nearest neighbors Regression (KNR), Linear Regression (LR), and Random Forest Regression (RFR), which similarly use the gene representations learned from BIONIC to predict transcriptional profiles. Network integration is the primary module for TranscriptionNet. An excellent network integration algorithm should produce accurate and comprehensive gene representations from biological networks. BIONIC used in our model has been proven to outperform existing integration methods across all evaluation types and benchmarks(17). To further confirm the potency of BIONIC in our experiment, we compare network integration results from BIONIC to three different established integration approaches: a naive union of networks (Union), a deep learning multi-modal autoencoder (deepNF)(18), and a multi-network extension of the node2vec(19) model (multi-node2vec)(20). The naive union of networks benchmark was created by taking the union of node sets and edge sets across input networks. For edges common to more than one network, the maximum weight was used. deepNF is an integrated framework based on a multimodal deep autoencoder that learns compact, low-dimensional feature representations of proteins from multiple heterogeneous interaction networks. It uses separate network layers to process different types of networks, fuses learned features into a bottleneck layer and finally performs SVM training on the resulting features(36). Multi-node2vec is an algorithm for multi-network extension of node2vec model, which learns node features from complex multi-layer networks through the Skip-gram neural network model[37]. For methods that produced features (deepNF, multi-node2vec, and BIONIC), a feature dimension of 512 was used to ensure results were comparable across methods. For methods that required a batch size parameter (deepNF and BIONIC), the batch size was set to 2,048 to ensure reasonable computation times. Except for the union method that has no hyperparameters, all other methods use their default hyperparameters. In addition, to ensure the advantage of gene representations encoded from multiple networks over single networks, we compare the network integration results of BIONIC with those using single networks. TranscriptionNet comprises two main stages, FunDNN and GenSAN. In addition, the second network GenSAN contains a "recycling" structure around the truck of the attention stack. We also examine the impact of the recycling process on the performance of TranscriptionNet.

**Coannotation analysis**

We evaluate the quality of predicted GECs by comparing their ability to predict the same functional terms in commonly annotated gene pairs between known GECs and predicted GECs. Here, we calculated the Pearson correlation coefficients between the target genes of known GECs (RNAi: 4454, CRISPR: 5139, OE: 3538) and predicted GECs (RNAi: 22496, CRISPR: 21806, OE: 23427). We obtained the relationship list between different annotation modules and target genes from Online Mendelian Inheritance in Man (OMIM)(24)、DrugBank、CORUM(25)、Pathway Reactome(26)、Kyoto Encyclopedia of Genes and Genomes (KEGG) pathways[21] and GO Biological Processes (BP)(22)databases. The two benchmarks of KEGG and GO Biological Processes (BP) were not used in our model as external benchmarks. For the GO Biological Processes (BP) benchmark, we removed annotation modules mapped to fewer than 20 and more than 500 target genes, leaving 4,203 annotation module-target gene relationship pairs. Comparing the annotation modules of two target genes, we take the two target genes with any common annotation module as the positive set, otherwise as the negative set. Considering the substantial imbalance between the positive set and the negative set, we extract the same number of target gene pairs from the negative set as the positive set, and perform five times of cross-validation. We used the area under the receiver operating characteristic curve (AUROC) and the area under the precision-recall curve (AUPRC) as quantitative standards to assess the ability of known and predicted GECs to predict coannotation genes, respectively.

**Characterization of compound-target interactions**

To investigate the relationship between drugs and protein targets, we downloaded LEVEL 5 from the CMap LINCS resource 2020 (version: beta) containing 720,216 GECs induced by small molecule compounds. We combine all the transcriptomic profiles of each compound using the weighted average algorithm in the L1000 project(13). Finally, we obtain unique transcriptomic profiles for 33,609 compounds.

We directly calculate the Pearson correlation coefficient between 33,609 compounds and 26,945 genetic perturbation GECs (including known GECs and predicted GECs), resulting in 100 million relationships. Among them, 10,000 pairs have been curated as true drug-target interactions by CMap. We refer to these interactions as the positive set, while keeping all other pairs as the negative set. First, for each type of genetic perturbation, we separately plot the absolute drug-target correlation distribution between the known and predicted GECs in the positive and negative sets. Additionally, to further quantify the performance of GECs in distinguishing different types of drug-target pairs, we calculate the true positive rate and false positive rate, and plot the ROC curves of known and predicted GECs based on different correlation thresholds. Finally, we summarize the correlation using logistic regression analysis for all three types of perturbations and plot the combined ROC curve of the known and predicted GECs.

**Characterization of disease-gene associations**

To investigate whether the predictive transcriptomic profile can effectively characterize the association between disease and gene, we collected transcriptomic data from the GEO database for human ischemic cardiomyopathy and non-ischemic cardiomyopathy (GEO database accession number: GSE46224)[38]. We use the R package DEseq2 (version: 3.18.0) to analyze the induced gene expression differences for ischemic cardiomyopathy and non-ischemic cardiomyopathy compared to the control group. We extract the 978 marker genes from the differentially expressed gene profiles for the two diseases.

We collected 110 genes associated with ischemic cardiomyopathy and 15 genes associated with non-ischemic cardiomyopathy from the current DisGeNET database (version: 7.0;

https://www.disgenet.org/)[39].

We directly calculate the Pearson correlation coefficients between 26,945 genetic perturbation GECs (including known GECs and predicted GECs) and marker genes in the differential expression profiles of diseases. We refer to the genes associated with diseases as the positive set, while all other genes were retained as the negative set. We randomly perform five runs. To quantify the performance of GECs in distinguishing disease genes, we use AUROC and AUPRC to evaluate the ability of known and predicted GECs to predict disease-gene associations, respectively.

**Data availability**

RNAi, OE, CRISPR, and compound data can be downloaded from the shared database at https://clue.io./releases/data-dashboard. Network data can be downloaded from the following databases: Disease (https://www.omim.org/), DrugBank (https://www.drugbank.ca/), CORUM (https://mips.helmholtz-muenchen.de/corum/), Pathway Reactome (https://reactome.org), STRING (https://string-db.org), UniProt (https://www.uniprot.org/), chromosomalLocation (http://www.ncbi.nlm.nih.gov/gene), KEGG (https://www.genome.jp/kegg/), GO_BP (https://www.geneontology.org/). Gene sets for ischemic cardiomyopathy and non-ischemic cardiomyopathy can be downloaded from DisGeNET (https://www.disgenet.org/). Transcriptional profiling data for ischemic cardiomyopathy and non-ischemic cardiomyopathy model groups can be downloaded from the Gene Expression Omnibus database with accession number GSE46224. In addition to the perturbations in the CMap dataset, GECs generated by RNAi, CRISPR, and OE from all gene members in the input network can be predicted through TranscriptionNet, which can be downloaded from https://github.com/lipi12q/TranscriptionNet/tree/master.

**Code availability**

The code for TranscriptionNet can be found at https://github.com/lipi12q/TranscriptionNet.